# Unveiling Saving and Credit Dynamics: Insights from Financial Diaries and Surveys among Low-Income Households in Unauthorized Colonies in Delhi.

Dr Divya Sharma Assistant Professor VIPS-TC, GGSIPU, Delhi &

PhD, Ambedkar University Delhi

divya.sharma@vips.edu Ph. No-9717759366

**Keywords:**
Financial Inclusion, Financial Diaries, Saving and Credit, Low-income households

## Abstract

The paper presents findings from a comprehensive study examining the saving and credit behaviors of low-income households residing in unauthorized colonies within a metropolitan area. Utilizing a dual approach, the study engaged in prolonged fieldwork, including repeated fortnightly interviews with selected households and a one-time primary survey with a larger sample size. The research meticulously analyzed the financial lives of these households, focusing on their saving and credit behaviors and assessing the accessibility and intensity of usage of financial instruments available to them. Through suitable regression models, the study identified key factors influencing the usage of financial instruments among low-income households. Transaction costs, convenience, and financial knowledge emerged as significant determinants impacting both usage decisions and the intensity of usage. The research underscores the importance of addressing demand side factors to ensure widespread financial services usage among low-income groups. Efforts to reduce time costs, enhance product accessibility and liquidity, and augment financial literacy are essential for fostering financial inclusion in unauthorized colonies. The findings highlight the imperative of moving beyond mere financial access towards promoting universal usage to realize the full benefits of financial inclusion.

## Introduction:

In the landscape of financial inclusion, understanding the saving and credit behavior of low-income households is paramount for policymakers and practitioners alike. This paper delves into the intricate dynamics of saving and credit decisions among low-income households residing in the three unauthorized colonies within a bustling metropolitan area. By employing a mixed-methods approach, which includes detailed analysis of financial diaries and comprehensive surveys, we aim to unravel the nuanced complexities underlying the financial lives of these households.

The methodology employed in this study is multifaceted, allowing for a holistic understanding of the saving and credit landscape within unauthorized colonies. We initiated our investigation with repeated fortnightly interviews, akin to financial diaries, with a carefully selected group of households. These interviews served the same purpose as financial diaries (used in papers such as Collins et al., 2009) and allowed us to gain the confidence of households and understand their financial lives closely. These repeated interviews gave us details regarding the day to day financial needs of the households. We gathered information about their assets, liabilities as well as their income and expenses. These interviews served as a window into the day-to-day financial activities of the participants, providing rich qualitative data that captured the intricacies of their saving and credit behaviors.

Complementing the insights gleaned from the financial diaries, we conducted a comprehensive one-time primary survey with a larger sample size. This survey enabled us to validate and augment our qualitative findings with quantitative data, offering a broader perspective on the saving and credit patterns prevalent among low-income households in unauthorized colonies. By triangulating data from both sources, we sought to enhance the robustness and depth of our analysis, thereby providing a comprehensive portrayal of the financial landscape under scrutiny.

Through our analysis, we aim to elucidate the factors driving saving and credit decisions among low-income households. From transaction costs and convenience to financial knowledge and household income, we explore the myriad influences shaping financial behaviors within this demographic. By uncovering these underlying determinants, we hope to offer actionable insights that can inform the design and implementation of tailored interventions aimed at bolstering financial inclusion among low-income communities.

We have studied a population in the metropolis where financial access is almost universal. This provides us an opportunity to look at factors other than the lack of supply which affects the usage of financial services and thus glimpse the challenges that lie ahead on the path of financial inclusion after access has been ensured for all. We look at low-income households who have unmet financial needs. Rutherford (1997) highlights that providing financial services to low-income households will help them manage their lives better even within the limited resources that they have.

For the low-income households providing financial access is necessary but not sufficient because they might face other obstacles like limited resources, low financial literacy and high transaction cost which can hinder usage. Even where supply of financial services exists, it is very much possible that the households decide not to save or borrow (Yaron,1992). It is possible that the households might just not have any surplus income (Iyer,2015) or find the cost of using these services to be too high. Low-income households in our study reside in unauthorised localities so they face the additional constraint of having only semi-legal titles to their houses, which means that the house owned cannot become a collateral for loans, as seen in the previous chapter.

We have chosen to study the financial lives of low-income households living in unauthorised localities as they are a relatively less studied segment of the Indian population when it comes to financial decision-making. We will show that financial access will make the household financially included but it does not automatically imply financial usage by them. Until and unless usage is universal, low-income households cannot reap the benefits of being part of the financial system. We will try to find out the possible factors which affect usage of financial services by evaluating the usage pattern as well as the intensity of usage of all the prominent financial instruments accessible to the households in these colonies under consideration.

In the subsequent sections of this paper, we delve into the findings derived from our analysis of saving and credit behaviors among low-income households in unauthorized colonies. Through a nuanced exploration of these dynamics, we endeavor to contribute to the ongoing discourse on financial inclusion, fostering greater understanding and action towards creating a more equitable and inclusive financial landscape for all.

**Related Work**

Karlan et al. (2014) argue that actual household savings fall short of expectations in a world with perfect competition and low transaction costs. The saving instruments offered to low-income households have several issues leading to under-savings. High transaction costs, such as account opening fees, travel expenses, and opportunity costs, discourage the use of financial products, especially among low-income households (Adams & Nehman, 2007). Additionally, formal financial institutions may intentionally maintain high transaction costs to deter low-income individuals (Karlan et al., 2014).

Trust issues also contribute to under-savings, as individuals may lack confidence in the security of their deposits and prefer to keep cash at home (Iyer, 2015; Dupas et al., 2012). Regulatory barriers further hinder access to banking services, particularly for the poor (Karlan et al., 2014).

Moreover, low financial literacy leaves many low-income households uninformed about available saving instruments (Dupas et al., 2012). This lack of knowledge exacerbates under-savings, as individuals may not understand the benefits of formal financial services.

Ravi and Gakhar (2015) identify prominent demand-side factors such as financial literacy, awareness, and lack of capacity, which perpetuate a vicious cycle of impoverishment. Financial capacity, determined by household income, significantly influences saving behavior, with low-income households facing limited saving capacity and difficulty accessing formal loans (Ravi & Gakhar, 2015).

On the supply side, high transaction costs deter the usage of financial services (Dupas et al., 2012). Despite initiatives like the Jan Dhan Yojana in India, low usage and account dormancy persist due to factors like fear of theft and service unreliability (Dupas et al., 2012).

## **Methodology used**

Our study belongs to the genre of literature that tries to study households' financial decisions through sustained engagements with them. We use financial diaries as a research methodology which is used to assess the household's earnings, spending as well as savings. It captures minutest details of information which are missed during snapshot interviews (RBI, 2013). It provides detailed and rich information about the financial behaviour of poor and low-income households. For this, repeated interactions were done on the same lines as financial diaries, with the only difference being that, here diaries are maintained by us and not by the respondents.

Some of the earliest research in the use of financial diaries to study the financial lives of the poor has been published in the book "Portfolios of the Poor: How the World's Poor Live on $2 a Day". Collins et al, 2009 mentions three studies in the book. The first study led by Stuart Rutherford in the year 1999 covered forty households in the slums of Bangladesh. Then a similar study took place in India by Orlanda Ruthven having close to fifty households. Next a study was conducted in urban, sub-urban and rural areas of South Africa with a larger sample of 152 households by Daryl Collins to collect information on cash flows daily. (Collins et al, 2009).

All the three studies mentioned in the book were year-long financial diaries of the slum dwellers and villagers in the three developing countries. Though one might have thought these low-income households would be living hand to mouth, immediately spending whatever they earn, but the reality brought out by these studies was that these households also do financial management and try to save for the bad times, emergencies and uncertainties using several financial tools, most of them through informal networks and connections. The book discusses the 'triple whammy' situation of poor and low-income groups: low and unpredictable income, inefficient financial instruments and the consequent poor financial management.(Collins et. al. 2009)

In 2013, the Reserve Bank of India (RBI) together with the Centre for Socio-Economic and Environmental Studies (CSES) undertook a one-month study using financial diaries for 107 below poverty line households in Kerala, whose daily cash flows were noted. The study found that on an average the cash inflow of the household was around fifteen thousand rupees per month, out of which sixty percent was from wages and salaries. The remaining forty percent was taken as loan. It is found that four out of five households had taken some loan to meet their shortfall. Many households took loans from their friends, relatives and neighbours. However all of these were small loans, which constituted only just one-fifth of the total outstanding loan amount. The most significant share of loan was taken from self-help groups followed by moneylenders (TNN. 2013)

Financial diaries go beyond one-shot surveys in capturing the minutest details of the cash flows, assets and liabilities of the households at a high-frequency in time. Any survey regarding financial details has to face the challenge of household's underreporting of information. Even the NSSO's AIDIS survey faces criticism on the grounds of underreporting of financial information (Satyasai. 2002). In the face of this limitation, the other large-scale Indian household survey, the CMIE's Consumer Pyramid Household Survey does not ask respondents for the exact value of financial assets and liabilities. Rather it has a ranking based system in which several preference options are given in the form of multiple-choice questions to have an idea about the financial lives of the respondents. Like for questions regarding the financial choices there are questions about how likely they are going to buy an insurance in future with options like most likely or unlikely.

The reluctance to share financial information was faced by us too when we began maintaining the financial diaries. There was hesitance to share true financial information with a complete stranger in the beginning because of the fear of misuse of information. Some households also feared that their information would be leaked to the income tax office. It is common to have such fears among the respondents when they meet for the first time and that is why one-shot surveys have an underreporting problem. But in the case of diaries since the selected households met regularly with the interviewer, they over time gained confidence regarding the safety of their information as well as the purpose of collection and as a result became more forthcoming

In the financial diaries approach the interviewer firstly briefs the respondent about himself/ herself, sets the intention clear and then meets on a repeated basis to establish a good rapport first. The scope of lies or underreporting also reduces because it's difficult to hide the true picture for a continuous period which is long enough. Another advantage is that the diaries are better in capturing the effects of shocks on households' financial lives as even the smallest impact can be noted which is possible to miss during a snapshot interview. In our case it could capture the impact of COVID pandemic during the course of the study.

To develop a qualitative analysis of low-income households in colonies we began with a series of structured interviews with a set of fourteen households. Table 1 below gives a brief description of the households selected for the financial diaries.

| Sno | Profession | Locality | Monthly Income | Gender | Age | Brief Note |
| --- | --- | --- | --- | --- | --- | --- |
| 1 | Peon | Budh Vihar | 10000 | Female | 52 | She is a peon of a private school gets Rs. 4500 and does socks packing in the evening. Family Income-Rs 10000 per month |
| 2 | Teacher | Budh Vihar | 10000 | Female | 35 | She has a fixed salary which is reduced from ten to six thousand in corona. Family Income-Rs 35000 per month |

| 3 | Cosmetic Shop Owner | Budh Vihar | 25000 | Female | 40 | She is a single mother and self-employed. Total Income- s 20-25000 per month |
| --- | --- | --- | --- | --- | --- | --- |
| 4 | Hospital Store In-charge | Budh Vihar | 15000 | Female | 37 | She is employed in a Private Hospital in Rama Vihar and earns 15000. Family Income- Rs 33000 per month |
| 5 | Electrician | Rama Vihar | 25000 | Male | 52 | He works as a free-lancer and earns twenty-twenty five thousand. Family Income- Rs 35000 per month |
| 6 | Mobile shop Owner | Rama Vihar | 30000 | Male | 45 | He is self-employed and doesn't employ any workers. Total Income- Rs 25-35000 per month |
| 7 | School Guard | Rama Vihar | 13000 | Male | 32 | He is employed in a private school, earning 13000. Family Income 20000 |
| 8 | Driver | Rama Vihar | 15000 | Male | 34 | He is employed by a wealthy property dealer, earning 15000. Family Income- Rs 15000 per month |
| 9 | Aganwadi Helper | Rama Vihar | 5000 | Female | 50 | She is an aganwadi helper and gets five thousand fixed salary<br>Family Income- Rs 25000 per month |
| 10 | Construction Contractor | Utsav Vihar | 25000 | Male | 54 | He hires five to seven workers when he gets a construction contract<br>Family Income- Rs 25000 per month |
| 11 | Momos Vendor | Utsav Vihar | 15000 | Male | 55 | He is self-employed and income reduced post corona from twenty to ten thousand<br>Family Income- Rs 15000 per month |
| 12 | Auto Driver | Utsav Vihar | Not working | Male | 54 | He had been unemployed during the pandemic and just resumed work in March.<br>Family Income- Rs 15000 per month |

| 13 | Statue Factory Owner | Utsav Vihar | 30000 | Male | 45 | He is self-employed and employees few workers<br>Total Income- Rs 30000 per month |
| 14 | Shop Worker | Utsav Vihar | 12000 | Male | 52 | He is employed in a private shop and gets twelve thousand<br>Family Income- Rs 12000 per month |

Table 1: Snapshot of Financial Diaries

The households were selected on the basis of convenience, and included a mix of self-employed and salaried workers. The interaction with these households spanned a period of seven months. A structured questionnaire was developed which had questions about their cash inflows and outflows as well as changes in assets and liabilities. For the first three months fortnightly personal interactions were held with respondents, which was followed by telephonic interviews once a month for the next four months. Each interview lasted for about fifteen to twenty minutes, where they were questioned according to the structured questionnaire as well as having a more open-ended discussion about their financial lives.

Almost all the households had migrated to the city at least ten to fifteen years ago., A few of them still had ancestral homes and some dependents living in the villages. Out of the fourteen respondents, eleven are from UP and others are from Rajasthan and Bihar. One respondent family hailed from Pakistan before the Partition, then they settled in some slums in Delhi and finally shifted to the colony. Several households have their friends and relatives from their villages who have also settled in the same colony.

## **Primary Survey**

Based on the inputs from these diaries, a questionnaire was developed for a larger survey which was carried out in August 2021. A total of two hundred households were selected through systematic sampling for interviews of about fifteen to twenty minutes. The questionnaire consisted of objective-type questions and have been attached in the appendix.

No sampling frame was available for these colonies. Due to the unauthorised nature of the colonies, many residents were still on the electoral rolls at the location of their ancestral homes. No comprehensive municipal records were available.

Under these conditions we used the following method for drawing the sample. Each of the three colonies have been divided into multiple blocks by the developers, labelled in alphabetic order from A to Z. Budh Vihar Phase-1 is the largest with twenty-five blocks, followed by Rama Vihar with six blocks and Utsav Vihar with only three.

We visited each street in each block of the three colonies. Blocks have been defined by the developers themselves for ease in administration. Every third house was approached for the survey. If the house was found locked or if the resident did not cooperate, we surveyed the next house. In a building with multiple households, one household was selected at random. Around a hundred households were surveyed from Budh Vihar Phase-1, followed by seventy-five from Rama Vihar and twenty-five from Utsav Vihar, roughly in proportion to the population of these colonies.

Budh Vihar Phase-1 has around 25- 30 streets and each street has been labelled as a block in alphabetic order from A to Z. Around hundred households have been surveyed from Budh Vihar Phase-1 with five or six households randomly picked up from each block. However, since few blocks mostly haveshops or mini factories, fewer households have been selected from those blocks and the shortfall has been made up from other blocks. The survey was carried out in Rama and Utsav Vihar in a similar manner.

## Descriptive Statistics

Table 2 summarises some of the basic characteristics of the survey respondents.

| Variables | BudhVihar | RamaVihar | UtsavVihar |
|---|---:|---:|---:|
| **Caste** | | | |
| General | 79 | 48 | 10 |
| OBC | 11 | 17 | 8 |
| SC | 9 | 8 | 7 |
| ST | 1 | 2 | 0 |
| Total | 100 | 75 | 25 |
| **Type of Employment** |  |  |  |
| Self Employed | 58 | 45 | 15 |
| Salaried | 26 | 22 | 4 |
| Others | 16 | 8 | 6 |
| Total | 100 | 75 | 25 |
| **Homeownership** |  |  |  |
| Owner | 76 | 47 | 16 |
| Rented | 24 | 28 | 9 |
| Total | 100 | 75 | 25 |

Table 2: Basic characteristics (counts)

Table 3 gives the mean and standard deviation of some key continuous variables.

| Variable | Budh Vihar | Rama Vihar | Utsav Vihar |
|---|---|---|---|
|  |  |  |  |
| Age of Household Head | **43.3** | **42.4** | **46.0** |
| | 12.2 | 9.9 | 7.8 |
| Size of the plot | **47.1** | **41.7** | **37.5** |
| | 39.1 | 39.7 | 38.0 |
| Household Size | **4.2** | **4.4** | **4.6** |
| | 1.3 | 1.2 | 1.3 |
| Non-working Members | **2.7** | **2.9** | **3.2** |
| | 1.3 | 1.4 | 1.4 |
| Total monthly family Income Pre-COVID (Rs) | **39300.0** | **36490.67** | **31700.0** |
| | 27983.2 | 24691.6 | 12293.3 |
| Total monthly family Income Post-COVID (Rs.) | **31500.0** | **29709.34** | **23240.0** |
| | 25278.3 | 43022.0 | 14360.0 |
| Cash at home (Rs.) | **10255** | **8401.3** | **2420.0** |
| | 16278.4 | 18333.8 | 1907.7 |
| Number of bank or ATM visits in past six months | **6.7** | **5.5** | **2.6** |
| | 5.1 | 4.0 | 3.1 |
| Value of Bank Deposits (Rs.) | **24867.0** | **25657.3** | **22196.0** |
| | 59923.0 | 65657.4 | 41546.8 |
| LIC coverage (Rs.) | **85900.0** | **155200.0** | **123000.0** |
| | 161500.8 | 232231.3 | 158429.8 |

Table 3: Means and Standard Deviations

The surveyed households primarily use two saving instruments: saving accounts and insurance, mostly life-insurance purchased from the LIC. We can see that despite the availability of banks, a good amount of cash is kept at home only by the households, which reflects a preference for holding cash for paying expenditures or emergencies as well as reflects low usage of saving accounts. It is seen that thirty-two per cent of total monthly family income is held as cash at home in Budh Vihar on an average, twenty-three per cent in Rama Vihar and roughly ten per cent in Utsav Vihar.

The average frequency of use of bank account/ATMs in the six months preceding the interview is between three to seven times. Roughly seven times in Budh Vihar meaning close to once a month, five times in Rama Vihar and only close to three times in Utsav Vihar.

If we look at the frequency of bank accounts/ATMs usage in the past six months at the time of the survey, we find out that thirty-five per cent of households have never used their bank account at all, so they remained dormant. Then forty-eight per cent of households have used it even less than one time a month, and sixty-six per cent used it once a month. Such once a month or even lesser can just be called usage for transactional purposes and not for saving purposes.

This observation is confirmed by our interaction with the branch manager of State Bank of India's Budh Vihar branch, according to whom close to seventy percent of accounts in that branch are pension accounts. Residents have to open accounts to receive their government transfer, cash subsidy or pension but they withdraw the money almost as soon as it is received. Just a small percentage of households use their bank accounts three to four times a month; We can count the use of an ATM as well as the bank account use together is considered as usage. With such a broad definition of frequency, using it three or four times a month is quite a low usage.

Next, coming to the usage of insurance policies, we find out that around fifty-five per cent of households have never bought any insurance for their family This is when all the three colonies have many LIC agents who are also familiar to everyone.

Thus, despite the availability of basic financial instruments which are even essential for low-income households, like saving accounts and insurance, the usage of both of them is not satisfactory so now we should explore the factors responsible for such low usage. In the rest of the chapter, we use our survey data and qualitative observations to try to understand this phenomenon of low usage by trying to determine which variables drive variations in usage among our study population.

Now we would use the regression analysis to understand the factors which affects usage of financial instruments. The factors affecting usage can be several which have been discussed in the literature review also like transaction cost, convenience, liquidity and literacy to name a few. We have quantified these variables and have been described in the next section. Let us look at the impact of each and every variable on the usage of financial instruments by these low-income households.

**Describing the variables**

Apart from the basic descriptive variables that we have defined in the previous section. Let us look at some of the other key categorical variables.

The first set of variables are related to transaction cost. Here we are using its proxy time cost because for the basic financial services that we are studying the pecuniary transactions costs are negligible and the primary component of cost is the time of the consumer. We measure it using the question: "how many visits are required for using the financial instrument or avail service?". It is a categorical variable that takes values: "do not know", "one visit" and "two or more visits".

Another proxy for transaction cost is how reachable the financial institution is from their residence. This varies among our respondents because except Budh Vihar, no other colony has a bank branch. The households have opened bank accounts in different bank branches in the surrounding authorised colonies and also some of the households have bank accounts in banks closer to their office or work area. Therefore, the next variable looks at the distance of bank branches or insurance agents from their homes.

We measure it using the question: "how far is the facility from which the financial instrument or service can be availed?" It is again a categorical variable that takes values: "walking distance", "cycle" or "rickshaw and bus or scooter".

While using any financial instrument households immediately look at the time involved in getting their money back. Is it available at demand or only at the time of maturity? Then even at maturity how quickly the money is received. As it is seen in the case of insurance policies, at the time of maturity some agents can quickly do all the paperwork and help you receive the sum assured very quickly, say within a week, for others it might be delayed and takes around a month. Same thing can happen when there is a claim If someone dies then how quickly they expect his/her dependents will get benefits differs among households, despite the financial instrument remaining the same.

The third variable, liquidity, is measured by responses to the question: "how much time does it take to withdraw the money deposited or invested in the financial instrument?" It is a categorical variable that takes values: "don't know", "one week" and "fifteen to thirty days".

The fourth variable, convenience, is constructed from responses to the question: "how easy it is to use the financial instrument or service?" It is a categorical variable that takes values: "difficult", "easy" and "very easy".

The fifth variable is the financial literacy ATM test, which is constructed by asking the respondents: "Do we need to put our account number while withdrawing money from an ATM?" From their responses a categorical variable was constructed that takes the value zero when the response is "do not know", and it takes value one when the response is "correct", and it takes value two when the response is "incorrect".

The sixth variable is again to test the respondents' financial knowledge of interest rates. The respondents were asked "Suppose you have rupees hundred in your savings account which earns two per cent rate of interest a year. After one year, how much would you have?" From their responses, a categorical variable was constructed that takes values: zero when the response is "do not know", one when the response is "correct", and it takes value two when the response is "incorrect".

The seventh variable influence is constructed from responses to the question: "Do you think your neighbours and friends are using a savings account?" It is a categorical variable that takes values: "not possible", "possible" and "highly possible".

## **Results**

Now before we delve into the regressions results we should discuss the motivation behind these set of regressions. We have focussed on two most used financial instruments by the households of these colonies that is bank account and LIC policies. The regressions have been aimed to understand the choices of financial mechanisms available to the low-income households under consideration. They are on the lines of the research questions which focuses on understanding the choices of financial instruments used by the households using the socio-economic characteristics of the household, social influence and transaction cost involved. Here we are using a logit regression to understand whether the financial instrument is used or not and then intensity of usage is analysed using a simple OLS regression.

We estimate the equation:

**Usage of bank account = Total family income post covid('000) + Age + Size of plot + time cost + reaching bank+ liquidity + convenience bank account+ convenience bank loan+ financial ATM test**

Average marginal effects                                                                 Number of obs =200

Model VCE: Robust

Expression: Pr (Usagebankac), predict()
Table 4: Marginal Effect of factors affecting usage of bank account

|  | dy/dx | std. err. | z | P>\|z\| | [95% conf. interval] | |
|---|---|---|---|---|---|---|
| | | Delta-method | | | | |
| Total family income | .001 | .001 | 1.650 | 0.098 | -.000 | .003 |
| age | -.004 | .002 | -1.950 | 0.051 | -.009 | .000 |
| Size of plot | -.000 | .000 | -0.090 | 0.928 | -.001 | .001 |
| Time cost | | | | | | |
| One Visit | .419 | .145 | 2.890 | 0.004 | .134 | .704 |
| 2-3 Visits | .300 | .142 | 2.11 | 0.035 | .021 | .580 |
| Reachability (Cycle/Rickshaw) Base: Walking distance | -.417 | .151 | -2.76 | 0.006 | -.714 | -.120 |
| Liquidity (1 week) Base: Don't know | .514 | .151 | 3.40 | 0.001 | .217 | .810 |
| Convenience | | | | | | |
| Easy | .079 | .061 | 1.280 | 0.199 | -.041 | .200 |
| Really Easy | .068 | .0894 | 0.77 | 0.443 | -.106 | .243 |
| Convenience bank loan | | | | | | |
| Easy | .151 | .049 | 3.060 | 0.002 | .054 | .249 |
| Really Easy | .022 | .094 | 0.240 | 0.813 | -.162 | .207 |
| Financial literacy ATM | | | | | | |
| Correct | -.037 | .056 | -0.67 | 0.504 | -.148 | .072 |
| Incorrect | -.065 | .081 | -0.81 | 0.420 | -.225 | .094 |

Note: dy/dx for factor levels is the discrete change from the base level.
Table 4: Marginal Effect of factors affecting usage of bank account

In Table 4, we find the non-users are kept as the base category. The result shows that the age of the household's head is significant and negatively affects being a user of a bank account. As the age increases by one unit, the probability of being a user of a bank account declines by 0.004 on an average. This could happen due to falling saving capacities as age increases, and even preference for holding cash is higher among the older generation.

Time cost involved in using the savings account also has a positive and significant effect. If the person can access its account in one go only without any hassle, then it induces usage. If the individual can access the savings account services in one go, then the probability of being a user will increase by 0.41 on an average. If it takes two to three times to access the bank account services, then the probability of being a user is lower; it will be 0.30 on an average.

Next, we see that as compared to walking distance to reach out for the banking service, if the distance increases and it takes a cycle or rickshaw, then this leads to 0.41 on an average reduction in being a user of the service.

After this, for low-income households, liquidity plays an important role in deciding to save in any instrument. From the don't know category, getting the financial asset liquidated within a week increases the probability of being a user of banking service by 0.51 on average.

Though convenience of using banking services is insignificant, convenience in getting a bank loan has a significant impact on usage. Now to avail any kind of bank loan, the minimum criteria is to have a bank account first.

If getting a bank loan becomes simpler, it would positively influence people to become a user of the banking services from being non-users. As we know, these households live in an unauthorised area, so their houses cannot act as collateral. Most of the households don't have enough resources to provide any other form of collateral. However, they do understand the value of a bank loan. That's why those households who possess ancestral property in their home town are used as collateral, and the loan is taken up. If required, the loan amount is used to meet the financial needs here.

This is so because bank loans seem convenient to them, and even interest rates are genuine. One of the dairy respondents, a mobile shop owner, had taken a loan from several places while constructing his house before the pandemic. His experience is worth noting as he is one of those who file ITR and also has the requisite documents to avail of a bank loan. He got a bank loan of around five lakh rupees from a private bank at a fourteen percent rate of interest. Then he took a similar amount of loan from a private finance company which charged him a much higher rate of interest, around 17-20 percent. He even took the remaining loan from a moneylender at a three percent monthly rate of interest, which comes out to roughly thirty-six percent annually.

Our next model looks at the intensity of usage among bank users by looking at the frequency of the bank account or ATM usage in the past six months at the time of the interview. We run a simple OLS regression, where the number of times a bank account is used (frequency of usage) is the dependent variable. It is a continuous variable, and it takes value starting from one.

**Frequency of Usage of bank account = Total family income post covid('000)+ house ownership+ time cost+ reaching bank+ liquidity bank+ convenience bank account + Influence convenience bank loan+ financial ATM test**

Linear regression                                         Number of obs    =    165
                                                          $F(12, 150)$    =    .
                                                          Prob > F        =    .
                                                          R-squared       =    0.3377
                                                          Root MSE        =    3.6433

**Linear regression**

| Frequency of Usage | Coef. | St.Err. | t-value | p-value | [95% Conf | Interval] | Sig |
|---|---|---|---|---|---|---|---|
| Total family income | .026 | .014 | 1.790 | .075 | -.003 | .054 | * |
| Homeownership: | 0 | . | . | . | . | . | |
| Rented | .305 | .637 | 0.48 | .633 | -.954 | 1.564 | |
| Transaction cost base (Don't know) | 0 | . | . | . | . | . | |
| Once | -1.114 | .986 | -1.130 | .260 | -3.061 | .834 | |
| 2-3 times | -.923 | 1.108 | -0.830 | .406 | -3.111 | 1.266 | |
| Reaching bank base (on foot) | 0 | . | . | . | . | . | |
| Cycle/rick | -3.212 | .823 | -3.90 | 0 | -4.837 | -1.587 | *** |
| liquidity base (Don't know) | 0 | . | . | . | . | . | |
| 1 week | .938 | .692 | 1.36 | .177 | -.429 | 2.306 | |
| Convenience base (Difficult) | 0 | . | . | . | . | . | |
| Easy | 1.162 | .652 | 1.78 | .077 | -.127 | 2.45 | * |
| Really Easy | 3.041 | 1.230 | 2.470 | .015 | .61 | 5.473 | ** |
| Influence base (Not Possible) | 0 | . | . | . | . | . | |
| Possible | -.189 | .908 | -0.21 | .836 | -1.983 | 1.606 | |
| Highly Possible | -1.660 | 1.109 | -1.50 | .136 | -3.851 | .531 | |
| Convenience bank loan base (Difficult) | 0 | . | . | . | . | . | |
| Easy | 1.694 | .682 | 2.48 | .014 | .346 | 3.042 | ** |
| Really Easy | 3.144 | 1.239 | 2.54 | .012 | .695 | 5.592 | ** |
| Financial literacy ATM base(Don't know) | 0 | . | . | . | . | . | |
| Correct | 1.888 | .591 | 3.20 | .002 | .721 | 3.055 | *** |
| Incorrect | -1.919 | .889 | -2.16 | .032 | -3.676 | -.163 | ** |
| Constant | 3.417 | 1.599 | 2.14 | .034 | .259 | 6.576 | ** |

| | | | |
|---|---|---|---|
| Mean dependent var | 6.933 | SD dependent var | 4.281 |
| R-squared | 0.338 | Number of obs | 165 |
| F-test | . | Prob > F | . |
| Akaike crit. (AIC) | 905.173 | Bayesian crit. (BIC) | 945.550 |

*** p<.01, ** p<.05, * p<.1

Table 5: Factors affecting frequency of usage of bank account

Here in Table 5, we see that the total family income of the household has a positive and significant effect on the frequency of usage. For a unit increase in total family income, the number of times a bank account is used increases by 0.26 on average, keeping everything else constant. This reflects that as income increases, then the frequency of usage increases.

The transaction cost comes out to be insignificant here which implies that it can play a limited role only as it can determine whether to use an instrument or not but it cannot decide the intensity of usage of bank account. Same goes for liquidity of the bank account.

Then we see that reachability of a bank branch or ATM also affects the usage positively and significantly. From being at walking distance, if the bank branch or ATM goes farther and it takes a cycle or rickshaw to travel there then keeping everything else constant, the frequency of usage declines by 3.22 on an average. This means that if the branch/ATM is at walking distance compared to taking a cycle or rickshaw to travel, then the frequency of usage increases.

The convenience of operating the bank accounts as well as for bank loans plays a positive and significant role in impacting the frequency of usage. Thus, it means that when financial instruments become easier to use, then the intensity of usage increases.

Next, we see that financial knowledge checked through the ATM test also positively and significantly impacts the frequency of usage. It is observed that if the ATM test question is correctly answered by the respondent, then the probability of becoming a user increases by 1.88 on an average compared to the situation when the respondent doesn't know it (base category)

Now there is another way by which we can see the intensity of usage that is through the amount of bank deposits kept by the account holders. To make it more specific, we take the amount of money saved in the bank as a fraction of total family income post covid, which becomes the dependent variable for this regression.

**Amount in bank/ total family income post covid = Total family income post covid('000)+ age+ size of plot+ not working+ house ownership + time cost+ liquidity + convenience bank +Influence + colony+ caste**

Linear regression                                        Number of obs   =   164
                                                         $F(16, 146)$    =   .
                                                         Prob > F        =   .
                                                         R-squared       =   0.1945
                                                         Root MSE        =   1.187

**Linear regression**

| bankamttoinc | Coef. | St.Err. | t-value | p-value | [95% Conf | Interval] | Sig |
|---|---|---|---|---|---|---|---|
| Total family income('000) | -.004 | .002 | -1.740 | .084 | -.008 | .001 | * |
| age | .001 | .008 | 0.180 | .859 | -.015 | .018 | |
| Size of plot | .012 | .005 | 2.55 | .012 | .003 | .021 | ** |
| Not working | .547 | .512 | 1.07 | .287 | -.465 | 1.559 | |
| Homeownership : base (owner) | 0 | . | . | . | . | . | |
| Rented | .451 | .329 | 1.37 | .173 | -.2 | 1.101 | |
| Time cost base (Don't know) | 0 | . | . | . | . | . | |
| Once | -.74 | .436 | -1.70 | .091 | -1.602 | .121 | * |
| 2-3 times | -1.22 | .452 | -2.70 | .008 | -2.115 | -.328 | *** |
| Liquidity base (Don't know) | 0 | . | . | . | . | . | |
| 1 week | .821 | .330 | 2.49 | .014 | .169 | 1.473 | ** |
| Convenience base (Difficult) | 0 | . | . | . | . | . | |
| Easy | .128 | .218 | 0.59 | .558 | -.303 | .559 | |
| Really Easy | .776 | .395 | 1.96 | .051 | -.005 | 1.556 | * |
| Influence Base (Not Possible) | 0 | . | . | . | . | . | |
| Possible | -.371 | .431 | -0.86 | .391 | -1.222 | .481 | |
| Highly Possible | -.413 | .534 | -0.77 | .441 | -1.469 | .643 | |
| Colony | 0 | . | . | . | . | . | |
| Base (Budh Vihar) | | | | | | | |
| RamaVihar | .107 | .176 | 0.61 | .543 | -.24 | .454 | |
| UtsavVihar | .423 | .411 | 1.03 | .305 | -.389 | 1.235 | |
| Caste | 0 | . | . | . | . | . | |
| base (General) | | | | | | | |
| OBC | .467 | .269 | 1.74 | .085 | -.065 | .998 | * |
| SC | .096 | .262 | 0.37 | .714 | -.422 | .614 | |
| ST | -.415 | .192 | -2.16 | .032 | -.794 | -.036 | ** |
| Constant | -.258 | .822 | -0.310 | .754 | -1.882 | 1.366 | |

| | | | |
|---|---|---|---|
| Mean dependent var | 0.818 | SD dependent var | 1.252 |
| R-squared | 0.194 | Number of obs | 164 |
| F-test | . | Prob > F | . |
| Akaike crit. (AIC) | 536.580 | Bayesian crit. (BIC) | 589.278 |

*** p<.01, ** p<.05, * p<.1

Table 6: Factors affecting amount of bank deposits

In Table 6, we find out that total family income is significant at one per cent, and it is negatively related to bank deposits made. For a unit increase in the total family income, the amount of deposits in the savings account as a fraction of total income declines by 0.04 on an average. This basically signifies that with higher income levels, the bank deposits as a proportion of income declines because households get other investment options. It no longer wants to restrict to a basic saving account.

The size of the plot of the household positively and significantly affects bank saving amounts. For a unit increase in the plot size of the household, the amount of deposit as a fraction of total income increases by 0.12 on an average.

Time cost again significantly affects saving deposits. A unit increase in time cost lowers the amount of savings deposited in a bank account by 0.74 on an average. However, when the transaction cost is even higher, it takes two to three visits for availing of banking services, then it declines the amount of money saved in bank accounts as a fraction of total income by 1.22 on an average.

Liquidity also plays a positive and significant role in the amount of savings done in the bank account as a fraction of total income. From being in the don't know category, if the money deposited in the bank account can be liquidated within a week, then the amount of money deposited as a fraction of total income increases by 0.82 on an average.

Convenience has a significant and positive role. For a unit increase in being really easy to operate a bank account from being difficult, it will positively influence the amount of money saved in bank account by 0.77 on an average.

In comparison to the general category, being in the OBC category will increase the amount of bank deposits by 0.46 on an average and being in the ST category will lower the amount saved in the bank as a fraction of total family income by 0.41 on an average. Both are significant.

Thus, we see that factors like transaction cost, convenience, liquidity, and reaching out to the bank significantly affect both usage that is deciding to be user or not and the intensity of usage. In comparison, the income of the household and financial knowledge affect the intensity of use only.

The next widely used financial instrument in our survey are life insurance policies. We run a simple logit model to evaluate the factors which decide whether insurance should be bought or not by the households. Here usage of insurance is the dependent variable which takes value one for being a user, and zero means Non-user.

**Usage of LIC= Transaction Cost + Reaching + Liquidity + Convenience Financial Literacy test**

Average marginal effects                     Number of obs = 200

Model VCE: Robust

Expression: Pr (Usagelic), predict()

|  | dy/dx | Delta-method std. err. | z | P>|z| | [95% conf. interval] | |
|---|---|---|---|---|---|---|
| Transaction cost(Don't know) | | | | | | |
| Once | .356 | .083 | 4.30 | 0.00 | .194 | .519 |
| 2-3 times | .285 | .181 | 1.57 | 0.116 | -.070 | .641 |
| Reaching out(on foot) | | | | | | |
| Cycle/rickshaw | .052 | .096 | 0.54 | 0.586 | -.136 | .240 |
| Bus/ Scooter | -.075 | .105 | -0.72 | 0.472 | -.281 | .130 |
| Liquidity (Don't know) | | | | | | |
| One week | .294 | .083 | 3.54 | 0.00 | .131 | .457 |
| 15-30 days | .240 | .096 | 2.48 | 0.013 | .050 | .430 |
| Convenience(difficult) | | | | | | |
| Easy | -.011 | .097 | -0.12 | 0.906 | -.201 | .178 |
| Really Easy | .214 | .111 | 1.92 | 0.055 | -.004 | .433 |

| Financial Literacy | | | | | | |
|---|---|---|---|---|---|---|
| Correct | .030 | .066 | 0.46 | 0.644 | -.099 | .161 |
| Incorrect | . | (not estimable) | | | | |

Note: dy/dx for factor levels is the discrete change from the base level.
Table 7: Marginal effects of factors affecting usage of LICs

In table 7, we find out that transaction cost has a significant effect. From being in the do not know category, if it takes just one visit for starting or renewing the policy, then the probability of becoming a user increases by 0.35 on an average. However, if it requires two to three visits for doing the same, then the probability of becoming a user declines by to 0.28 on an average. Thus, a lower transaction cost increases the probability of being a user significantly.

After that, we see that the liquidity of LICs also has a significant effect on being a user. It is observed that from the do not know category moving to be able to get money back within a week, the probability of being a user increases by 0.29 on an average. While moving from the do not know category to get money back within fifteen to thirty days positively increases the probability of being a user by 0.23 on an average. Thus, lowering the liquidity time period will significantly increase the probability of being a user.

When it comes to liquidity, then there is another interesting benefit that the policyholders can avail themselves closer to maturity. That is, they can take a loan against the policy from the LIC itself in case of emergencies. One such loan against the policy from the LIC is taken by an auto driver, who is also one of the diary respondents. He had to get his daughter married and was in need of urgent money. He approached the LIC agent and was suggested to take a loan against the matured sum due next year on an interest close to one percent monthly.

Then we move to the intensity of usage of the service, which is determined by the average amount of LIC cover taken by the household. To be specific, we take a ratio of amount of LIC cover divided by the total pre covid income of the household. This ratio basically tells the number of months of insurance the household has.

**LIC Coverage/total family income pre covid = Total family income pre covid('000)+ age+ size of plot + Transaction cost + Reachability + Liquidity + Convenience**

Linear regression                                                       Number of obs   =   90

|  |  |  |  |
|---|---|---|---|
| F(9, 80) | = | 2.20 |
| Prob > F | = | 0.0304 |
| R-squared | = | 0.2650 |
| Root MSE | = | 10.334 |

**Linear regression**

| LICcovertoinc | Coef. | St.Err. | t-value | p-value | [95% Conf | Interval] | Sig |
|---|---|---|---|---|---|---|---|
| Total Family income | -.132 | .048 | -2.760 | .007 | -.227 | -.037 | *** |
| age | -.025 | .135 | -0.180 | .855 | -.294 | .244 |  |
| Size of the plot | .031 | .032 | 0.970 | .334 | -.032 | .094 |  |
| Transaction cost | .89 | 1.521 | 0.580 | .56 | -2.135 | 3.914 |  |
| Reachability | -.182 | 1.889 | -0.100 | .923 | -3.94 | 3.575 |  |
| Liquidity | -1.01 | 2.317 | -0.440 | .664 | -5.619 | 3.6 |  |
| Convenience | -1.464 | 2.01 | -0.730 | .468 | -5.463 | 2.534 |  |
| Constant | 16.038 | 5.365 | 2.99 | .004 | 5.365 | 26.712 | *** |

| Mean dependent var | 8.031 | SD dependent var | 11.428 |
|---|---|---|---|
| R-squared | 0.137 | Number of obs | 90 |
| F-test | 1.210 | Prob > F | 0.307 |

| | |
| --- | --- |
| Akaike crit. (AIC) | 695.621 |
| Bayesian crit. (BIC) | 715.619 |

*** p<.01, ** p<.05, * p<.1

Table 8: Factors affecting number of months of insurance taken by the household

In table 8 we mention the results of a simple OLS regression where the dependent variable is the number of months for which households are insured, that is the insurance coverage taken by the household as a fraction of total family income. We see that the total family income of the household has a significant effect, and it is negatively related here. That is, for a unit increase in income, the number of years of insurance taken by the household declined by 0.025. This could be possible due to the fact that when total family income increases, then the number of options for investment increases and the household will like to move beyond insurance.

All in all, we have seen that when it comes to deciding to be a user of a financial instrument then time cost, liquidity, convenience, and reaching out play a significant role. However, in case of bank account, when it comes to the intensity of usage, we find that the financial capacity, that is, the household's total income affects the frequency of usage and amount of deposits differently. It is seen that income of the household and frequency of usage are positively related. This means that when income increases, the number of times banking services are used within a given time frame of study increases. But when it comes to the amount of deposits made as a fraction of total family income, then it is observed that an increase in income would lead to a decline in deposits made. This could mean that the household would get other investment options with a rise in income levels and would like to move beyond the bank's savings.

Regarding the LIC policies, we find out that transaction cost and liquidity play an important role in determining whether the household would be a user or not. Financial knowledge about the LIC policy and basic interest rates awareness also positively affects usage.
However, when we look at the intensity of usage, then the average amount of LIC coverage taken by the family as a fraction of total family income has a negative relation with total income. This means that higher-income households like to move beyond insurance.

## Conclusion:

We find that despite broad availability of financial services, there is large variation in their usage. Our key findings suggest that transaction cost, convenience and financial knowledge play a significant role in deciding whether the financial service is used or not as well as in the intensity of usage. On the other hand, the household's income has a significant effect on the intensity of usage alone and not on usage versus non-usage decisions. Thus focus on increasing financial capacity of household will augment the intensity of usage of financial services.

Ensuring usage among these low-income groups depends on several demand and supply-side factors. This study has highlighted some key factors like financial knowledge, financial capacity, and availability of suitable financial products that induce usage among households. Efforts are required to reduce the time costs, offer products with a high degree of liquidity, and augment the financial literacy among low-income households to reap the true potential of financial inclusion.

Looking forward, there remains ample scope for further research in this area. Future studies could delve deeper into specific aspects of saving and credit behavior, such as the impact of digital financial services, the role of social networks, or the effectiveness of financial education programs. Additionally, longitudinal studies could provide valuable insights into the dynamics of financial behavior over time and the effectiveness of interventions aimed at promoting financial inclusion.

Moreover, expanding the scope of research to include comparative studies across different geographic regions or demographic groups could offer valuable insights into the factors driving saving and credit behaviours in diverse contexts. Furthermore, qualitative research methods, such as in-depth interviews or focus group discussions, could provide deeper insights into the motivations and decision-making processes of low-income households.

In conclusion, our study lays the groundwork for further research and underscores the importance of continued efforts to promote financial inclusion among low-income communities. By building on these insights and exploring new avenues of inquiry, we can work towards creating a more equitable and inclusive financial landscape that empowers all individuals to achieve their financial goals and aspirations.

# Appendix
## QUESTIONNAIRE

# Informality and financial decisions of the low-income groups

## Personal Characteristics

**Name**
___________________________

**Mobile Number**
___________________________

**Address**
___________________________

**Caste**
- ○ General
- ○ General-EWS
- ○ OBC
- ○ SC
- ○ ST

## Family Details

**Name of the member**
___________________________

**Gender**
- ☐ Male
- ☐ Female
- ☐ Transgender

**Age**
___________________________

**Relation to Respondent**
- ○ Father
- ○ Mother
- ○ Grand Father
- ○ Grand Mother
- ○ Son
- ○ Daughter
- ○ Grand daughter
- ○ Grandson
- ○ Daughter in law
- ○ son in law
- ○ wife
- ○ brother
- ○ sister
- ○ self

**Occupation**

___________________________________

**Nature of Compensation**
- ○ Fixed Salary/Wage
- ○ Commission
- ○ Self Employed

**Employment Status**
- ○ Employed in private sector
- ○ Employed in public sector
- ○ Unemployed
- ○ Student
- ○ Retired

**If unemployed, where where you employed earlier or pre covid?**

___________________________________

Income earned pre covid

Income earned post covid

If Job, then please share Provident fund and pension details if you get them?

## Income details

Household income from rent?

Total Family income pre Covid

Total Family Income post covid

Household income from pension and cash transfers

## Mobiles owned

Number of mobiles

Financed by

Asset value

Asset Age

## Refrigerator owned

number of Refrigerator

Financed by

Asset Value

Asset Age

## AC owned

number of AC

Financed by

Asset Value

Asset Age

___________

## Coolers owned

number of coolers

___________

Financed by

___________

Asset Value

___________

Asset Age

___________

## Washing Machine owned

number of washing machine

___________

Financed by

___________

Asset Value

___________

Asset Age

___________

## TV Owned

**number of TV**

**Financed by**

**Asset Value**

**Asset Age**

## Computer Owned

**number of computer**

**Financed by**

**Asset Value**

**Asset Age**

## Car Owned

number of car

[           ] ⇅

Car type

______________________________

Financed by

______________________________

Asset Value

[           ] ⇅

Asset Age

[           ] ⇅

## Two Wheeler Owned

number of Two wheelers

[           ] ⇅

Type of Two wheeler

______________________________

Financed by

______________________________

Asset Value

[           ] ⇅

Asset Age

[           ] ⇅

## Bank Saving Account

**Frequency of Usage of bank accounts**
*How many times do you save in them during the past 6 months?*

_______________

**Accumulated Amount**

_______________

## Fixed Deposit

**Frequency of Usage of Fixed deposits**
*How many times do you save in them during the past 6 months?*

_______________

**Accumulated Amount**

_______________

## Shares and Mutual Funds

**Frequency of Usage of shares & mutual funds**
*How many times do you save in them during the past 6 months?*

_______________

**Accumulated Amount**

_______________

## Saved in Business

**Frequency of Usage of saved in business**
*How many times do you save in them during the past 6 months?*

**Accumulated Amount**

## Gold and Silver

**Frequency of Usage of gold and silver**
*How many times do you save in them during the past 6 months?*

**Accumulated Amount**

## Real Estate

**Frequency of Usage of real estate**
*How many times do you save in them during the past 6 months?*

**Accumulated Amount**

## Cash at Home

**Frequency of Usage of cash at home**
*How many times do you save in them during the past 6 months?*

**Accumulated Amount**

_____________________

## Committees

**Frequency of Usage of committees**
*How many times do you save in them during the past 6 months?*

_____________________

**Accumulated Amount**

_____________________

## Saved with Friend and Relative

**Frequency of Usage of saved with friend**
*How many times do you save in them during the past 6 months?*

_____________________

**Accumulated Amount**
*How many times do you save in them during the past 6 months?*

_____________________

## LIC and Health Insurance

**When did you start?**

_____________________

**Coverage**
*If a person dies then what would they get*

[          ⇕ ]

**Premium paid**
*Annually*

[          ⇕ ]

**Maturity**

[          ⇕ ]

| Saving Potential | Not Possible | Possible | Easily Possible | |
|---|---|---|---|---|
| How easy it is for you to save at least one tenth from your earnings every month? | ○ | ○ | ○ | |
| How often do you have to borrow for regular monthly expenses? | ○ | ○ | ○ | |
| **Transaction Cost** | Once | 2-3 Times | More than 4 | Don't Know |
| How many times you need to visit the bank for opening a saving account? | ○ | ○ | ○ | ○ |
| How many times you need to go to open a Fixed Deposit? | ○ | ○ | ○ | ○ |
| How many times you need to go to open a Post Office account? | ○ | ○ | ○ | ○ |
| How many times you need to go to start a LIC policy? | ○ | ○ | ○ | ○ |
| **Transaction Cost** | On Foot | Cycling/Rickshaw | Bus/Scooter/Car | |
| How can you reach a bank branch/ATM to deposit money? | ○ | ○ | ○ | |
| How can you reach a post office to deposit money? | ○ | ○ | ○ | |
| How can you reach a LIC agent to give premium? | ○ | ○ | ○ | |
| How can you reach a Committee organizer for depositing money? | ○ | ○ | ○ | |
| **Liquidity** *How many days will it take to get back money from these saving instruments?* | 1 week | 15-30 days | More than a month | Don't Know |
| Bank Saving Account | ○ | ○ | ○ | ○ |
| Fixed Deposit | ○ | ○ | ○ | ○ |
| Post Office Savings | ○ | ○ | ○ | ○ |
| Kisan Vikas Patra | ○ | ○ | ○ | ○ |
| Provident Funds | ○ | ○ | ○ | ○ |
| LIC | ○ | ○ | ○ | ○ |
| Shares and Mutual Funds | ○ | ○ | ○ | ○ |
| Saved in Business | ○ | ○ | ○ | ○ |
| Gold & Silver | ○ | ○ | ○ | ○ |
| Real Estate | ○ | ○ | ○ | ○ |
| Committees | ○ | ○ | ○ | ○ |

| Convenience | Difficult | Easy | Really Easy |
|---|---|---|---|
| How easy it is to deposit money in saving account? | ○ | ○ | ○ |
| How easy it is to deposit money in Fixed Deposit? | ○ | ○ | ○ |
| How easy it is to deposit money in Post Office? | ○ | ○ | ○ |
| How easy it is to have a LIC policy? | ○ | ○ | ○ |
| How easy it is to start a Committee? | ○ | ○ | ○ |

| Homology | Not Possible | Possible | Highly Possible |
|---|---|---|---|
| Do you think your neighbours and friends like to save at home only? | ○ | ○ | ○ |
| Do you think your neighbours and friends have a saving account? | ○ | ○ | ○ |
| Do you think your neighbours and friends have a fixed deposit? | ○ | ○ | ○ |
| Do you think your neighbours and friends have a LIC policy? | ○ | ○ | ○ |
| Do you think your neighbours and friends deposit money in committees? | ○ | ○ | ○ |

| Social Influence | Not important | Important | Very important |
|---|---|---|---|
| *How important are the following to take guidance regarding major financial decisions?* | | | |
| Elders in the family | ○ | ○ | ○ |

Purpose
_________________________

Source
_________________________

Tenure
_________________________

EMI paid (monthly)
_________________________

Interest
_________________________

Type of interest
○ Monthly
○ Quarterly
○ Yearly

| Transaction cost for loans | Once | 2-3times | More than 4 | Don't Know |
|---|---|---|---|---|
| How many times you need to visit the relative/friend before you get the borrowing | ○ | ○ | ○ | ○ |
| How many times you need to visit the affluent acquaintance before you get the borrowing? | ○ | ○ | ○ | ○ |
| How many times you need to visit the moneylender before you get the borrowing? | ○ | ○ | ○ | ○ |
| How many times you need to visit the private company before you get the borrowing | ○ | ○ | ○ | ○ |

| Access for Interest Free Loans | Difficult | Easy | Really easy |
|---|---|---|---|
| Do you think everyone can get small amount of borrowings from relatives and friends | ○ | ○ | ○ |
| Knowledgable Neighbours | ○ | ○ | ○ |
| Friends | ○ | ○ | ○ |
| Colleagues and Employer | ○ | ○ | ○ |
| Media Advertisments | ○ | ○ | ○ |

| Savings Pattern | Not Confident | Confident | Very Confident |
|---|---|---|---|
| Have you been able to save for short term contingencies? | ○ | ○ | ○ |
| Have you been able to save money for an accident/job loss? | ○ | ○ | ○ |
| Have you been able to save for unpredictable health loss or loss of life of the bread earner? | ○ | ○ | ○ |
| Have you been able to save for your retirement? | ○ | ○ | ○ |
| Have you been able to save for your children's Marriage | ○ | ○ | ○ |
| Have you been able to save for children's education | ○ | ○ | ○ |

Have you liquidated any asset post covid?

___________________________________________

| Demand for Liquidity | Never | Sometimes | Regularly |
|---|---|---|---|
| How often does unexpected emergencies arise? | ○ | ○ | ○ |
| How often does the cash at home is insufficient to meet the expenses? | ○ | ○ | ○ |
| How often does the cash at home is insufficient to meet the unexpected expenses? | ○ | ○ | ○ |
| How often does unpredictable events like job loss struck the household? | ○ | ○ | ○ |
| How often does major health issue struck the household? | ○ | ○ | ○ |

## Inventory of Loans

Amount

______________

https://ee.kobotoolbox.org/preview/jRzIvCEZ8    Page 13 of 20

If there is urgent need of money amounting above 2 lakhs rupees, who among the following would you approach?

- [ ] Bank
- [ ] Private Company
- [ ] Money Lender
- [ ] LIC
- [ ] Property Dealers
- [ ] Others

## For House Owners

**Size of the plot**

____________________

**House type- Number of rooms**

____________________

**When was this constructed?**

____________________

**When was the last major construction done?**

yyyy-mm-dd

**Do you give some rooms on rent?**

- ( ) yes
- ( ) No

**If yes, What is the rent charged?**

____________________

**Source of finance**

____________________

|  |  |  |  |
|---|---|---|---|
| How easy it is to approach a private company for loan? | ○ | ○ | ○ |
| How easy it is to approach a moneylender for loan? | ○ | ○ | ○ |

## Collateral

Is there any collateral asked by bank for the loan? If yes, please specify?
______________________________________________

Is there any collateral asked by private company for the loan? If yes, please specify?
______________________________________________

Is there any collateral asked by moneylender for the loan? If yes, please specify?
______________________________________________

If there is urgent need of money amounting between 10-40000 rupees, who among the following would you approach?

- ☐ Bank
- ☐ Private Company
- ☐ Money Lender
- ☐ Employer
- ☐ Affluent People
- ☐ Property Dealer
- ☐ relative
- ☐ friends

If there is urgent need of money amounting between 50000-2 lakhs rupees, who among the following would you approach?

- ☐ Bank
- ☐ Private Company
- ☐ Money lender
- ☐ LIC
- ☐ Property Dealers
- ☐ Others

How optimistic are you that you would never be evicted from here?
- ○ Very optimistic
- ○ optimistic
- ○ Fear eviction

## Source of Investment (Ask home owners only)

Number of loans taken where house was kept as collateral

_______________

Did you get a loan because you own a house with semi-legal title?
- ○ Yes
- ○ No

Can you easily find buyers for your home, if you wish to sell it?
- ○ Yes
- ○ No
- ○ Dont know

Can you easily get the cash in hand incase you sell it in emergencies?
- ○ Yes
- ○ NO
- ○ Dont Know

Do you think the price of the plot have increased since you have purchased?
- ○ Yes
- ○ No
- ○ Dont Know

Can you easily take a loan from an informal institution using house as a collateral?
- ○ Yes
- ○ No
- ○ Dont know

## on rent

**Are you planning to buy your own home in next 5 years?**
- ○ Yes
- ○ No
- ○ Not Decided

**When did you last change your home?**

yyyy-mm-dd

**Do you like to buy house in the current locality you are living?**
- ○ Yes
- ○ No
- ○ Not decided

**What is the rent paid?**

| Financial | Regularly | In some months &seasons | Never |
|---|---|---|---|
| How often are you surprised where is your money gone at the end of the month? | ○ | ○ | ○ |
| Do you make a financial plan? | ○ | ○ | ○ |
| Did you keep worrying about your financial situation pre covid? | ○ | ○ | ○ |
| Do you keep worrying about your financial situation post covid? | ○ | ○ | ○ |
| For frequent worriers | Lose sleep | Worse Health | Less productive at work |
| For worriers | ○ | ○ | ○ |

**Do we need to put our account number while withdrawing money from ATM?**
- ○ Yes
- ○ No
- ○ Dont Know

**Suppose you have Rs 100 in your saving account which earns 2 percent rate of interest a year. After 5 years, how much would you have?**

- ○ More than 102
- ○ Exactly 102
- ○ More than 110
- ○ Don't Know

**A 15 year loan typically requires higher monthly payments than a 30 year loan, but the total interest paid over the life of the loan will be less?**

- ○ True
- ○ False
- ○ Dont Know